\documentclass[
 reprint,
superscriptaddress,
bibnotes,
showkeys,
amsmath,amssymb,
aps,prb,
floatfix
]{revtex4-1}

\usepackage{graphicx}
\usepackage{dcolumn}
\usepackage{bm}
\usepackage{natbib}
\usepackage[margins]{trackchanges} 

\addeditor{KAM} 

\begin{document}

\title{Observation of the signatures of sub-resolution defects in two-dimensional\\superconductors with scanning SQUID}

\author{Hilary Noad}
\affiliation{Stanford Institute for Materials and Energy Sciences, SLAC National Accelerator Laboratory, 2575 Sand Hill Road, Menlo Park, CA 94025, USA}
\affiliation{Department of Applied Physics, Stanford University, Stanford, CA 94305, USA}
\author{Christopher A. Watson}
\affiliation{Stanford Institute for Materials and Energy Sciences, SLAC National Accelerator Laboratory, 2575 Sand Hill Road, Menlo Park, CA 94025, USA}
\affiliation{Department of Applied Physics, Stanford University, Stanford, CA 94305, USA}
\author{Hisashi Inoue}
\affiliation{Department of Applied Physics, Stanford University, Stanford, CA 94305, USA}
\author{Minu Kim}
\affiliation{Stanford Institute for Materials and Energy Sciences, SLAC National Accelerator Laboratory, 2575 Sand Hill Road, Menlo Park, CA 94025, USA}
\author{Hiroki K. Sato}
\affiliation{Stanford Institute for Materials and Energy Sciences, SLAC National Accelerator Laboratory, 2575 Sand Hill Road, Menlo Park, CA 94025, USA}
\author{Christopher Bell}
\affiliation{Stanford Institute for Materials and Energy Sciences, SLAC National Accelerator Laboratory, 2575 Sand Hill Road, Menlo Park, CA 94025, USA}
\affiliation{HH Wills Physics Laboratory, University of Bristol, Tyndall Avenue, Bristol, BS8 1TL, United Kingdom}
\author{Harold Y. Hwang}
\affiliation{Stanford Institute for Materials and Energy Sciences, SLAC National Accelerator Laboratory, 2575 Sand Hill Road, Menlo Park, CA 94025, USA}
\affiliation{Department of Applied Physics, Stanford University, Stanford, CA 94305, USA}
\author{John R. Kirtley}
\affiliation{Geballe Laboratory for Advanced Materials, Stanford University, Stanford, CA 94305, USA}
\author{Kathryn A. Moler}
\affiliation{Stanford Institute for Materials and Energy Sciences, SLAC National Accelerator Laboratory, 2575 Sand Hill Road, Menlo Park, CA 94025, USA}
\affiliation{Department of Applied Physics, Stanford University, Stanford, CA 94305, USA}
\affiliation{Geballe Laboratory for Advanced Materials, Stanford University, Stanford, CA 94305, USA}

\begin{abstract}

The diamagnetic susceptibility of a superconductor is directly related to its superfluid density. Mutual inductance is a highly sensitive method for characterizing thin films; however, in traditional mutual inductance measurements, the measured response is a non-trivial average over the area of the mutual inductance coils, which are typically of millimeter size. Here we measure localized, isolated features in the diamagnetic susceptibility of Nb superconducting thin films  with lithographically defined through holes, $\delta$-doped SrTiO$_3$, and the 2D electron system at the interface between LaAlO$_3$ and SrTiO$_3$, using scanning superconducting quantum interference device susceptometry, with spatial resolution as fine as 0.7 $\mu$m. We show that these features can be modeled as locally suppressed superfluid density, with a single parameter that characterizes the strength of each feature. This method provides a systematic means of finding and quantifying submicron defects in two-dimensional superconductors. 
\end{abstract}
\maketitle

\section{Introduction}
Two-dimensional superconductors, in which the superconducting thickness, $d$, is much smaller than the London penetration depth, $\lambda$,\cite{PearlAPL64} are both technologically important and central to active fields of research in condensed matter physics. Thin film cuprates \cite{BozovicNat16} make it possible to explore fundamental properties of the cuprate phase diagram. Superconducting complex oxide heterostructures, such as LaAlO$_3$/SrTiO$_3$ (LAO/STO) \cite{OhtomoNat04, ReyrenSci07} and $\delta$-doped SrTiO$_3$ (STO), \cite{KozukaNat09} exhibit tunable two-dimensional superconductivity. Exquisite control of heterostructure growth enables engineered systems demonstrating high-temperature superconductivity.\cite{ge2018superconductivity} Wafer-scale superconducting electronics rely on well-controlled thin films.\cite{tolpygo2016advanced}
Both fundamental studies and development of applications of superconductivity require an understanding of, if not complete control over, typical defects that may influence or obscure the intrinsic effects of interest. 

In this paper, we image the susceptibility of thin-film samples using a Superconducting QUantum Interference Device (SQUID) susceptometer with two concentric micron-scale loops. We find approximately circular features (``halos'') in susceptibility with a minimum at the center and an asymmetry that mimics our sensor layout.  We interpret these features as regions of reduced superfluid density that are small compared to the size of our sensors ($\sim 1~\mu$m). Sub-resolution defects have the simplest possible geometry for a defect in two dimensions, as they are effectively point-like for the purposes of our measurements. Some possible intrinsic sources of such defects could include the intersection of a line defect, such as a crystallographic dislocation, with the superconducting plane; small patches of phase-separated material; or dopant inhomogeneity. Sub-resolution defects can also be intentionally added to a system, for example, in ion irradiation experiments that test the sensitivity of the superconductivity to changes in scattering or to create vortex pinning sites in order to improve the critical current of superconducting wires. 

Two-coil mutual inductance experiments with millimeter-scale spatial resolution have been an excellent method for characterizing the area-averaged properties of thin-film superconductors for many years.\cite{FioryAPL1988penetration,ULMPRB1995magnetic} Our susceptometers \cite{HuberRevSciInstrum08, KirtleyRevSciInstrum16} have two micron scale pickup loops integrated into two-junction scanning Superconducting Quantum Interference Devices (SQUIDs) in a gradiometric configuration. In addition, each pickup loop is paired with a one-turn, co-planar, concentric field coil. This provides a similar geometry to the two-coil mutual inductance experiments mentioned above, but with better spatial resolution and high sensitivity even at low, quasi-DC frequencies. Here we demonstrate how these SQUID susceptometers detect defects in two-dimensional superconductors that are much smaller than the length scales of our sensor. We first describe a simple model \cite{KoganPRB11} and compare it to results for artificial defects with variable sizes. We then present experimental results on two 2-dimensional systems and find that the model reproduces the effect that we observe in real samples under reasonable assumptions for sample and imaging parameters. 

\section{Experimental Methods}

\subsection{Data acquisition}
We imaged the susceptibility of a Nb film with intentionally introduced holes and in two two-dimensional superconductors using scanning SQUID microscopes in a liquid helium cryostat (Nb films) and in a dilution refrigerator \cite{bjornsson2001scanning}($\delta$-doped STO, LAO/STO). The Nb films were $0.2~\mu$m thick sputtered films containing lithographically defined through holes. The $\delta$-doped STO and LAO/STO were grown by pulsed laser deposition; details of the growth of the $\delta$-doped STO can be found in Ref. \onlinecite{NoadPRB16} (sample with doped layer $5.5$ nm thick, 1 at.\% Nb doping) and of the LAO/STO in Ref. \onlinecite{BertPRB12}.  Details of the imaging conditions (temperatures, field coil current, and excitation frequency) are noted in the figure captions. The SQUID sensors used in this study were Nb-Al$_2$O$_3$-Nb trilayer susceptometers.\cite{HuberRevSciInstrum08,KirtleyRevSciInstrum16} We define the susceptibility, $\phi=\Phi/I\Phi_0$, as the mutual inductance of the SQUID field coil/pickup loop pair in the presence of the sample. This mutual inductance is given by the ratio of the flux, $\Phi$, through the SQUID pickup loop to the current through the field coil, $I$, normalized by the superconducting flux quantum $\Phi_0=h/2e$.  $\phi$ has two components: $\phi_1$ is the flux response in phase with the field coil current, acquired by recording the flux signal with zero phase shift relative to the field coil current using a lock-in amplifier; $\phi_2$ is the out-of-phase response, with the lock-in set to a 90 degree phase shift relative to the field coil current. A negative value for $\phi_1$ corresponds to a diamagnetic susceptibility, while a positive value corresponds to a paramagnetic susceptibility.

\subsection{Image processing}
For the images of holes in thin films of niobium 
in Fig. \ref{fig:multiple_experiment}, 
we subtracted the susceptibility at a corner of each image, to more directly compare the data with Ref. \onlinecite{KoganPRB11}, which calculates to first order in a perturbation expansion. For the images plotted in Figs.~\ref{fig:examples}(a), \ref{fig:comparison}(a), and \ref{fig:tser_dSTO}(a, b), we measured the susceptibility as a function of height $z_0$ (see Fig. \ref{fig:halo_geometry}), recording both quadratures with an SR830 lock-in amplifier, before taking each susceptibility image. We also recorded both quadratures of the susceptibility signal during imaging. To set the zero in these images, we subtracted the average of the 20 points furthest from the sample in susceptibility vs. height for each quadrature independently. 
For the in-phase susceptibility images of LAO/STO plotted in Fig.~\ref{fig:examples}(b), Fig.~\ref{fig:tser_LAOSTO}(a), and Fig.~\ref{fig:FCser_LAOSTO}(a), we subtracted a susceptibility offset following the same procedure as described above for the $\delta$-doped STO data; for the out-of-phase data, for the temperature series, we subtracted the mean of the image taken at the highest temperature, Fig.~\ref{fig:tser_LAOSTO}(b)(vii), and for the field coil series, the mean of the image taken at the lowest field coil current, Fig.~\ref{fig:FCser_LAOSTO}(b)(i).

\section{\label{sec:methods_modeling}Theoretical Modeling}
\subsection{In-phase susceptibility}
\label{sec:kogantheory}
\begin{figure}
\centering
\includegraphics[width=\linewidth]{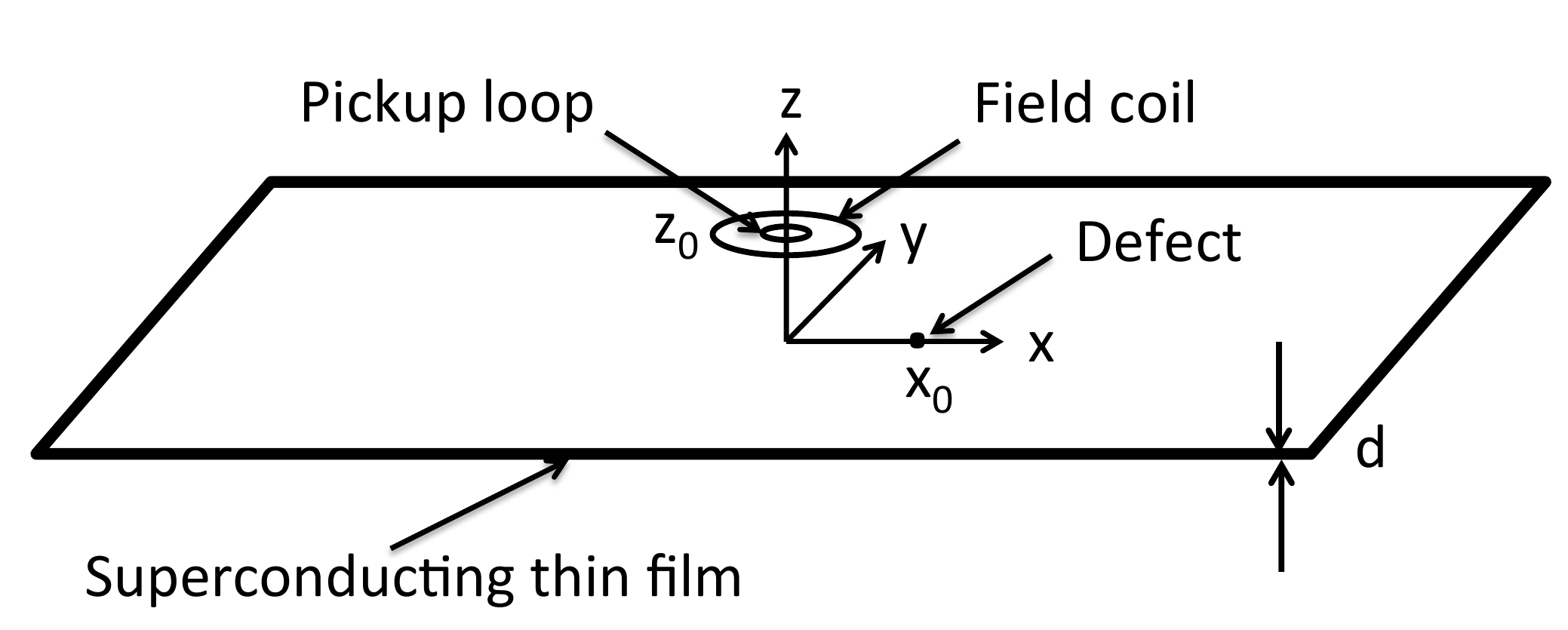}
\caption{Assumed geometry. A superconducting film of thickness $d$ and infinite extent in the $x$ and $y$ directions is centered on the plane $z=0$. The susceptometer field coil and pickup loop are modeled as co-planar, infinitely thin circular loops of radius $a$ and $b$ respectively, oriented parallel to the thin film in the plane $z=z_0$. A point defect is at $x=x_0, y=0, z=0$.}
\label{fig:halo_geometry}
\end{figure}
We used the approach outlined in Ref.~\onlinecite{KoganPRB11} to calculate the in-phase susceptibility signal due to the reduction or enhancement of the Pearl length associated with defects 
as imaged by a scanning SQUID susceptometer microscope. For the sake of completeness we summarize here the results of this approach for a point defect in a thin film. 

Our assumed geometry is illustrated in Figure \ref{fig:halo_geometry}. 
Several assumptions go into the calculation.

First, we assume that the superconducting film is in the Pearl limit, \cite{PearlAPL64} with the unperturbed London penetration depth, ${\lambda}_{0}$, larger than the superconducting film thickness, $d$. In this limit, the characteristic magnetic length scale is the Pearl length, ${\Lambda}_{0} = 2{\lambda}_{0}^{2}/d$. We further assume that the dimensions of the region of modified penetration depth are small compared to all other length scales of the problem, namely ${\Lambda}_{0}$, the radius of SQUID field coil, $a$, and the radius of SQUID pickup loop, $b$.
In modeling the SQUID itself, we use the approximation that the field coil and pickup loop are each infinitely thin, perfectly circular, and continuous loops, are exactly co-planar and concentric with each other, and are exactly parallel to the superconducting plane.

The first assumption is not met for our experiments with holes in Nb, since their thickness (0.2$\mu$m) is several penetration depths ($\lambda_0 \approx 0.08 \mu$m) thick. These experiments do however support our interpretation of haloes as being due to local reductions in the superfluid density that are on a length scale shorter than the dimensions of our susceptometer pickup loops. The first assumption is, however, comfortably met for the $\delta$-doped STO sample ($\Lambda_0$ $\approx$ 2 mm) and the LAO/STO sample ($\Lambda_0$ $\approx$ 25 mm) as indicated by fits to our data. Separate analyses of vortices\cite{BertThesis12} indicate Pearl lengths of many hundreds of $\mu$m or longer in the $\delta$-doped STO sample. 

We do not know the underlying physical origin(s) of the defects that produce halos in the susceptibility of either the $\delta$-doped STO or the LAO/STO, so we do not know precisely how well the second assumption is satisfied. However, if the defects were similar in size to the pickup loop of the SQUID, we would expect to begin to see substructure or other deviations from the model in the measured susceptibility profile. We do not see such deviations for the naturally occurring defects studied here. Likewise, for the intentionally introduced defects in Nb films (Fig. \ref{fig:multiple_experiment}), the image of the smallest hole appears similar to the naturally occurring halos.  However, as the holes become larger, the minimum in susceptibility at the center of the defect image gradually disappears.
We will cover the applicability of the final assumptions in Sec.~\ref{ch_pointlike:sec:results}, in our discussion of the results.

In the absence of vortices London's equation for the fields inside a superconductor is written (in S.I. units) as
\begin{equation}
\vec{h}+\vec{\nabla}\times (\lambda^2 \vec{j}) = 0,
\label{eq:london}
\end{equation}
where $\vec{h}$ is the magnetic field, $\lambda$ is the (in our case inhomogeneous) London penetration depth, and $\vec{j}$ is the superfluid density. Writing $\vec{h}=\vec{h}^s+\vec{h}^r$, where $\vec{h}^s$ is the field from the field coil, and $\vec{h}^r$ is the response field from the superconductor, and integrating over the film thickness, results in the following for the $z$ component of the field:
\begin{equation}
h_z^s+h_z^r+\frac{\Lambda}{2} \left ( \frac{\partial g_y}{\partial x} - \frac{\partial g_x}{\partial y} \right ) +\frac{1}{2} \left( g_y \frac{\partial \Lambda}{\partial x} -g_x \frac{\partial \Lambda}{\partial y} \right ) =0,
\label{eq:2dLondon}
\end{equation}
where $\Lambda=2\lambda^2/d$ is the Pearl length, and $g_x$ and $g_y$ are the 2-d supercurrent densities in the $x$ and $y$ directions respectively. Integrating Maxwell's equation $\vec{\nabla}\times \vec{h}=\vec{j}$ over the film thickness, and neglecting terms proportional to $\partial h_z/\partial y$ and $\partial h_z/\partial x$, results in 
\begin{eqnarray}
g_x&=&-2h^r_y(0^+) \nonumber \\ 
g_y&=&2h^r_x(0^+),
\label{eq:gfromh}
\end{eqnarray}
where the response fields are evaluated at the top surface of the thin film. Then Eq. \ref{eq:2dLondon} becomes, also at the top surface but suppressing the $0^+$ notation: 
\begin{equation}
h_z^s+h_z^r+\Lambda \left ( \frac{\partial h_x^r}{\partial x}+\frac{\partial h_y^r}{\partial y} \right ) + \left ( h_x^r \frac{\partial \Lambda}{\partial x}+ h_y^r \frac{\partial \Lambda}{\partial y} \right ) = 0
\label{eq:2dLondon2}
\end{equation}
We model the defect as a point deviation in the Pearl length: 
\begin{equation}
\Lambda(\vec{r}) = \Lambda_0-\gamma^3\delta(\vec{r}-\vec{r}_0),
\label{eq:GammaLambda}
\end{equation}
where $\gamma$, with the dimensions of a length, represents the strength of the defect. We can write the field as the gradient of a scalar potential: $\vec{h} = \vec{\nabla}\varphi$. Expanding the source  and response scalar potentials in the region of space  $0<z<z_0$ in Fourier series 
\begin{eqnarray}
\varphi^s(\vec{r},z) &=& \frac{1}{2\pi^2} \int_0^\infty  \varphi^s(\vec{k})e^{i\vec{k}\cdot\vec{r}+kz} d^2k  \nonumber \\
\varphi^r(\vec{r},z) &=& \frac{1}{2\pi^2} \int_0^\infty  \varphi^r(\vec{k})e^{i\vec{k}\cdot\vec{r}-kz} d^2k
\label{eq:varphi}
\end{eqnarray}
results in
\begin{equation}
\varphi^r(\vec{k})=\frac{\varphi^s(\vec{k})}{1+\Lambda_0k}+\delta\varphi^r(\vec{k})
\label{eq:varphi2}
\end{equation}
with
\begin{equation}
\delta\varphi(\vec{k}) = \frac{\gamma^3e^{-i k_x x_0}}{(2\pi)^2 k(1+k\Lambda_0)} \int \varphi^s(\vec{q}) \frac{\vec{k}\cdot\vec{q}e^{i q_x x_0}}{1+q\Lambda_0} d^2q
\label{eq:dvarphi}
\end{equation}
The first term in Eq. \ref{eq:varphi2} is the standard result for the susceptibility of a homogeneous thin superconductor;\cite{KoganPRB11} the second term is the change in the response due to inhomogeneity.  The source potential for a circular loop of radius $a$ carrying current $I$  is given by\cite{kogan2003meissner}
\begin{equation}
\varphi^s(\vec{k}) = \frac{\pi I a}{k} e^{-k z_0} J_1(ka),
\label{eq:phisource}
\end{equation}
where $J_1$ is the Bessel function of the first kind of order 1.

Substituting Eq. \ref{eq:phisource} into Eq. \ref{eq:dvarphi} results in
\begin{eqnarray}
h^r_z(\vec{k})&=&-k\,\delta\varphi(\vec{k}) \nonumber \\
&=&-\frac{\gamma^3Iae^{-ik_xx_0}}{4\pi(1+k\Lambda_0)}\int \frac{\vec{k}\cdot\vec{q}J_1(qa)e^{iq_xx_0-qz_0}}{q(1+q\Lambda_0)}d^2q
\label{eq:hzk}
\end{eqnarray}
for the $z$-component of the response field. Fourier transforming Eq. \ref{eq:hzk}, integrating over the pickup loop area $S$, and using the relations $\int_S e^{i\vec{k}\cdot\vec{r}} d^2r = 2\pi bJ_1(kb)/k$ and $\int_0^{2\pi} \cos(\theta) e^{\pm i z \cos(\theta)} d\theta = \pm 2\pi i J_1(z)$ results for the in-phase response flux through the pickup loop
\begin{equation}
\phi^r_1 =\frac{\Phi}{I\Phi_0}
= -\frac{\mu_0\gamma^3ab}{2\Phi_0}D_x(a)D_x(b)
\label{eq:finalphi}
\end{equation}
where $\Phi$ is the flux through the pickup loop, $\Phi_0=h/2e$ is the superconducting flux quantum, and
\begin{equation}
D_x(r)=\int_0^\infty \frac{q J_1(qr)J_1(qx_0)e^{-qz_0}}{(1+q\Lambda_0)}dq
\label{eq:Dx}
\end{equation}

\begin{figure}
\centering
\includegraphics[width=\linewidth]{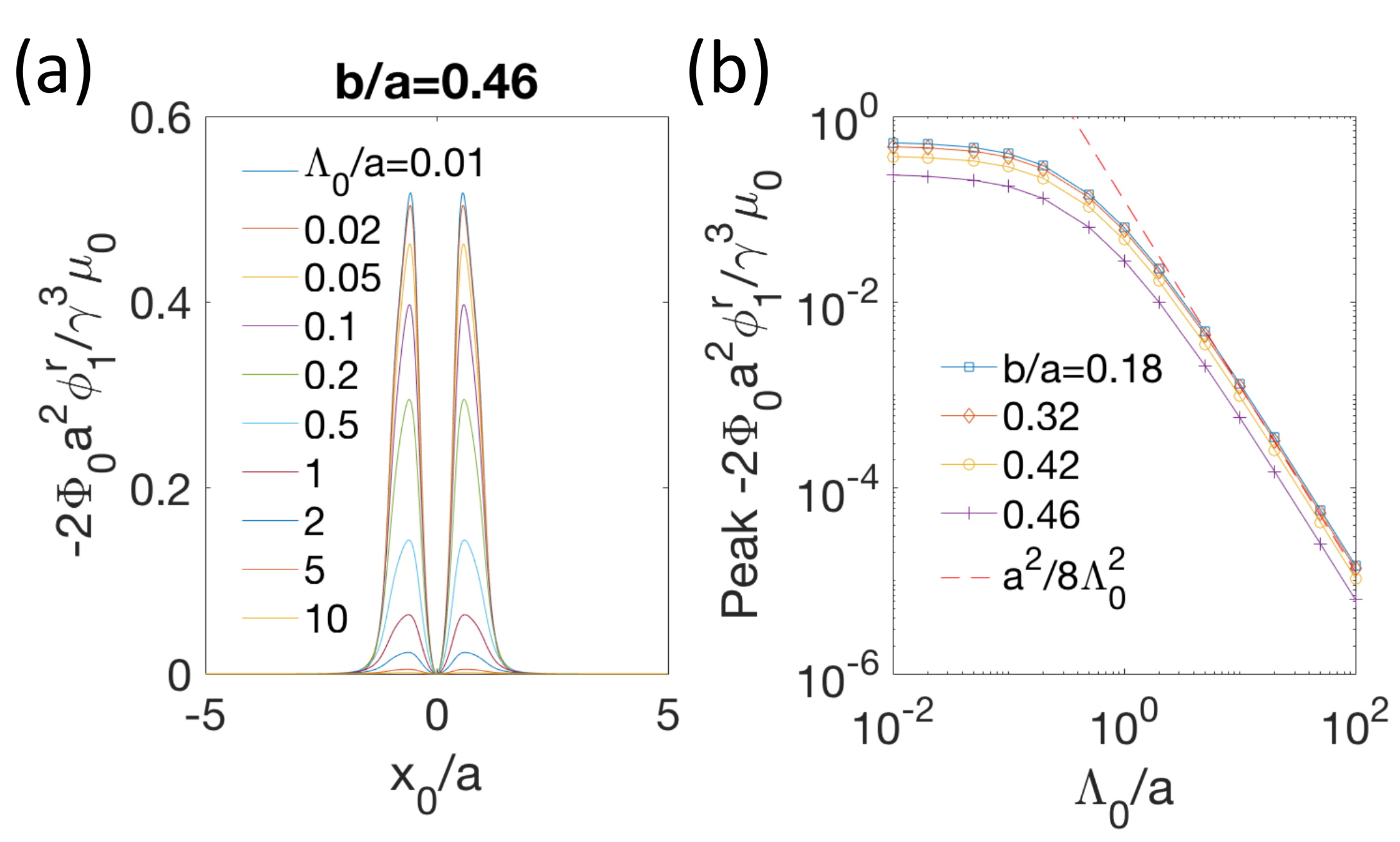}
\caption{{\bf  In-phase susceptibility of defects in thin film superconductors. } (a) Plot of normalized in-phase defect response susceptibility $-2\Phi_0a^2\phi_{r,1}/\gamma^3\mu_0$ vs lateral spacing $x_0/a$, for various reduced Pearl lengths $\Lambda_0/a$, for spacing $z_0/a=0.25$ between the sample surface and the field coil/pickup loop (see Fig. \ref{fig:halo_geometry}), and $b/a$=0.46. (b) Peak value of the normalized in-phase defect response susceptibility $\phi_1^r$ as a function of $\Lambda_0/a$, for various values of $b/a$, with $z/a$=0.25. For large values of the Pearl length $\phi_1^r$ is approximately proportional to $1/\Lambda_0^2$; for small $\Lambda_0$'s $\phi_1^r$ is nearly proportional to $1/a^2$. In all cases the defect response susceptibility scales with the cube of the defect strength parameter $\gamma$. }
\label{fig:plot_haloes}
\end{figure}

Figure \ref{fig:plot_haloes}a plots the predictions of Eq. \ref{eq:finalphi} for the in-phase response flux as a function of lateral spacing $x_0$ for a number of values of $\Lambda_0/a$, for fixed $z_0/a$=0.25 and $b/a$=0.46. Note that the response flux goes to zero as $x_0$ goes to zero. This is because when the field coil is centered directly above the defect, there are no circulating Meissner shielding currents at the defect. The response curve peaks at $x_0 \approx a$: the halos are about the same size as the field coil used to image them. Further, the response flux becomes weaker as the homogeneous Pearl length $\Lambda_0$ becomes larger. Figure \ref{fig:plot_haloes}b plots the peak value of $\phi^r_1$ as a function of $\Lambda_0/a$ for a number of values of $b/a$. 

\subsection{Out-of-phase susceptibility}
We will now turn to the interpretation of the out-of-phase susceptibility signal that we see in LAO/STO, but not in $\delta$-doped STO, at the same location as the in-phase signal.

It is important to first rule out spurious effects that could mimic physically interesting ones.  The simplest way to get an unphysical out-of-phase signal is to have an unaccounted-for $RLC$ time constant coming from somewhere in the experimental setup. In such a scenario, a signal that starts out purely in-phase acquires a phase shift 
\begin{equation}
{\theta}_{RLC} = {\tan}^{-1}\left (\frac{{\omega}L-\frac{1}{{\omega}C}}{R}\right),
\end{equation}
where $R$, $L$, and $C$ are the parasitic resistance, inductance, and capacitance of the experimental setup, and $\omega$ is the frequency of the signal. Crucially, the phase shift depends on frequency but \emph{not} on the amplitude of the signal. 

All of the images presented below for LAO/STO 
were taken at the same frequency, $1863.3$ Hz. If the phase were due to RLC effects, we would expect it to be constant within a given scan and from scan to scan. As such, while these effects may give a small, uniform phase offset, they cannot account for the strong dependence of the phase on position and field coil amplitude.  We therefore interpret the out-of-phase signal in the LAO/STO data
as coming from physical, dissipative processes in the sample, as described below. 

Mutual inductance measurements of the kind we consider here effectively measure a complex susceptibility, $\chi = \chi_1+i\chi_2$, where the real part describes the superfluid response while the imaginary part is related to dissipation and power loss.\cite{ClemBook}  In measurements of the mutual inductance of superconductors, the imaginary part has been shown to peak near $T_c$,\cite{Hebard79} and has also been used to obtain information on such physics as inter- and intra-granular critical fields in granular superconductors,\cite{SchulzSSC91, SinghPRB97} vortex states and dynamics,\cite{JonssonPRB98} and the spatial distribution of superconductivity, e.g. whether it is filamentary or not.\cite{MaxwellPRL63}

As has been demonstrated previously,\cite{[{See for example: }] Hebard82,*Jeanneret89,*Clem92,*Turneaure98, Hebard79, HahnThesis} the complex mutual inductance can be numerically related to the complex conductivity, $G$, which is further the inverse of the complex impedance $Z = 1/G = R + i\omega L$, where $R$ is a generalized resistance and $L$ is a generalized inductance.  In the zero field limit, where vortex contributions to the impedance can be taken to be real, the inductive part of the complex impedance is given by the kinetic inductance of the superfluid,\cite{HahnThesis} from which the Pearl length and, by extension, the superfluid density and London penetration depth, can be extracted. 

While the geometric factors involved in extracting these superconducting parameters from scanning SQUID susceptometry measurements have been considered previously,\cite{KirtleyPRB12} extracting the complete complex conductivity in our geometry has not. By obtaining the complex conductivity, we could, in principle, extract from the out-of-phase halo signal a local reduction of the critical current density near the observed defects with numerical modeling.\cite{ClemSanchez}

\section{\label{ch_pointlike:sec:results}Results}
\subsection{\label{intentional}Intentionally produced defects}
\begin{figure}
\centering
\includegraphics[width=\linewidth]{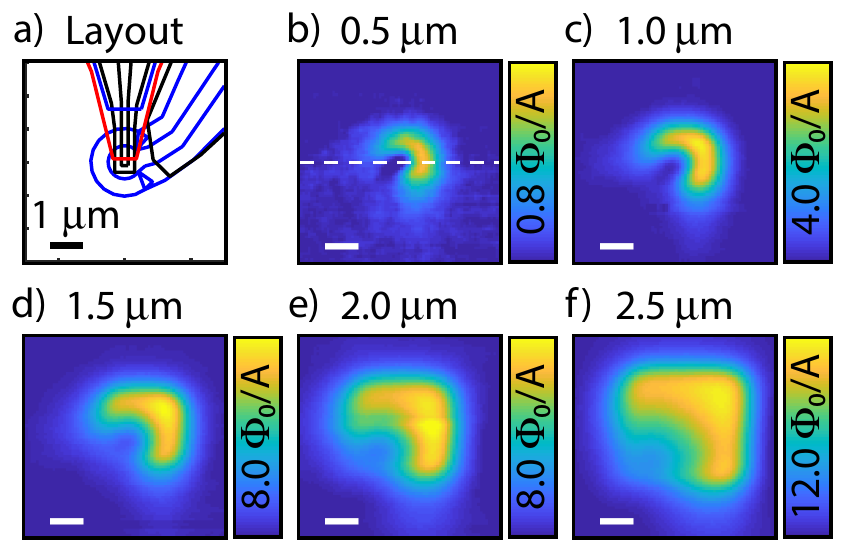}
\caption[]{{\bf Experimental susceptibilities of lithographically defined holes through a 0.2 $\mu m$ niobium film.} (a) Pickup loop (black)/field coil (blue) layout for the susceptometer used. The red layer is a superconducting shield for the pickup loop. (b)-(d) Susceptibility images for square holes with sizes as labeled. Field coil current 1 mA at 2.024 kHz, $T$ = 5 K. The SQUID substrate and sample were touching during the scan, such that $z_0$, the spacing between the sample surface and the pickup loop layer, was about 0.5 $\mu$m. The dashed line in (b) shows the location of the cross-section through the data displayed in Fig. \ref{fig:multiple_theory}.}
\label{fig:multiple_experiment}
\end{figure}
\begin{figure}
\centering
\includegraphics[width=\linewidth]{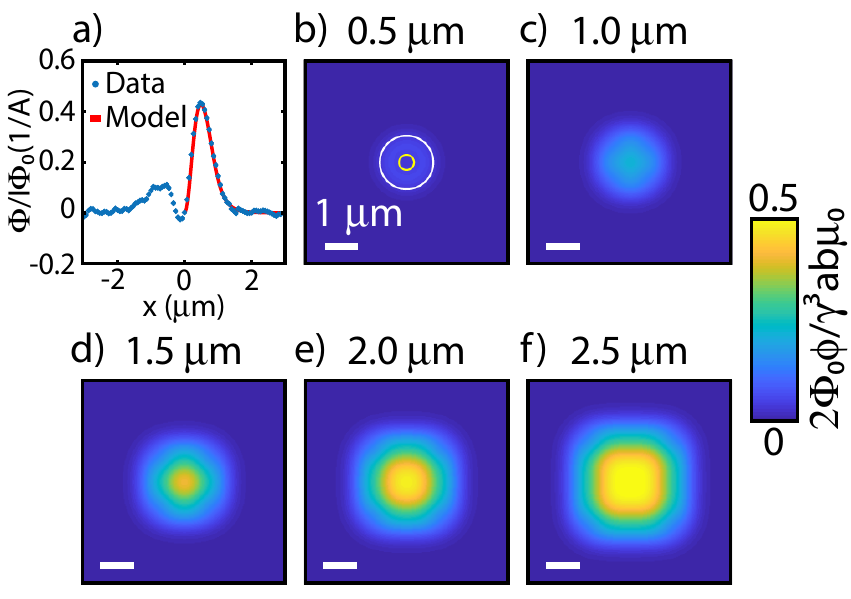}
\caption[]{{\bf Theoretically predicted susceptibilities.} (a) The symbols are a cross-section as indicated in Fig. \ref{fig:multiple_experiment}(b). The solid line is from the  model\cite{KoganPRB11} described above (Eq.~\ref{eq:finalphi}), with $b$= 0.79 $\mu m$ (white circle)  and  $a$=0.22 $\mu m$ (yellow circle) in Fig. \ref{fig:multiple_theory}(b). The fitting parameter $\gamma$ = -0.5 $\mu$m. (b)-(d) Calculated susceptibilities for square holes in a niobium film  with the sizes as labeled, obtained by convolving the point spread function from (b) with the hole shapes.}
\label{fig:multiple_theory}
\end{figure}

\begin{figure}
\centering
\includegraphics[width=\linewidth]{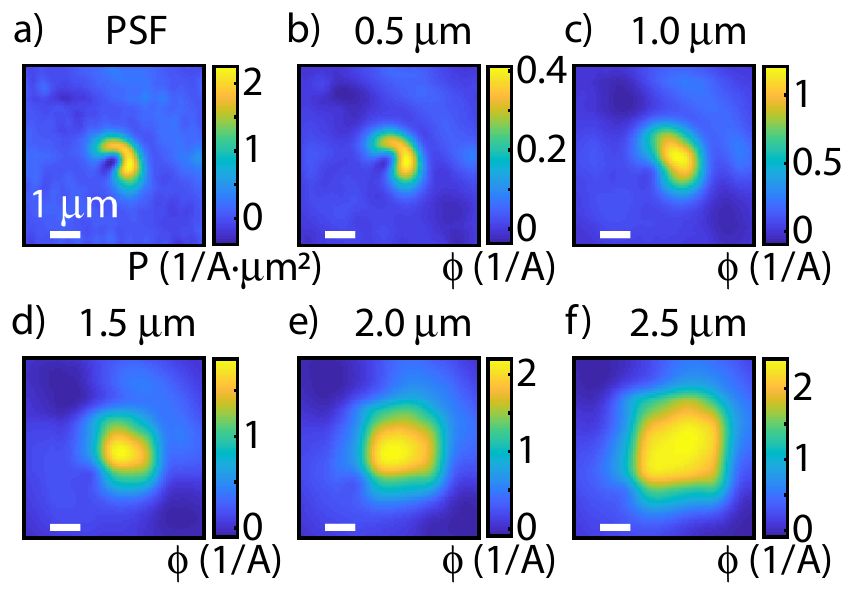}
\caption[]{{\bf Calculated susceptibilities using an experimental point spread function.} (a) Point spread function inferred from experimental susceptibility of a 0.5 $\mu m \times 0.5~\mu m$ hole in niobium. (b)-(d) Calculated susceptibilities for rectangular holes with sizes as labeled.}
\label{fig:multiple_psf}
\end{figure}

As a test of the theory  described in Section \ref{sec:kogantheory}\cite{KoganPRB11} we imaged the susceptibility of lithographically fabricated holes in a 0.2 $\mu$m thick Nb film. This theory can only be expected to apply qualitatively to these experiments, since it is developed in the Pearl limit $\lambda  > d$, while the sample had $\lambda$ ($\approx$ 0.07$\mu$m) $<$ d (0.2 $\mu m)$. For these experiments, we used a susceptometer with the highest spatial resolution available, with 0.2 $\mu$m inside diameter pickup loops,\cite{KirtleyRevSciInstrum16} as diagrammed in Fig. \ref{fig:multiple_experiment}(a). In this layout, the base electrode, which constitutes the field coil and the lower shield for the pickup loop, is in blue; the first wiring level, which constitutes the pickup loop, is in black; and the second wiring layer, which constitutes the upper shield for the pickup loop, is in red. The panels in Fig. \ref{fig:multiple_experiment}(b-d) are susceptibility measurements for square holes of different sizes, as labeled. The similarity of these measurements to the naturally occurring halos described below supports our interpretation of such features as being due to defects in the superconducting films.

The symbols in Fig. \ref{fig:multiple_theory}(a) represent a cross-section as indicated by the dashed line through the data of Fig. \ref{fig:multiple_experiment}(b). The solid line in Fig. \ref{fig:multiple_theory}(a) is a a fit to Eq. \ref{eq:finalphi}, 
with the fitting parameter $\gamma = -0.5~\mu$m chosen to fit the data on the right hand side of the cross-section. The effective field coil radius $b=\sqrt{(r_{in}^2+r_{out}^2)/2}=0.79~\mu m$, where $r_{in}$ is the inner radius and $r_{out}$ is the outer radius of the field coil. The effective pickup loop radius $a=0.22~\mu m$ was chosen to match the measured mutual inductance between the field coil and the pickup loop.\cite{KirtleyRevSciInstrum16}
Figure \ref{fig:multiple_theory}(b) shows the predicted susceptibility for a point defect with these parameters.

Modeling of the SQUID susceptibility signal from defects of finite size is difficult. One possible approach is to sum up the contribution from multiple small defects separated in space from one another. The panels in Fig. \ref{fig:multiple_theory}(c-f) are calculated susceptibility images using
\begin{equation}
\phi(x,y) = \int_{-\infty}^{\infty} dx' \int_{-\infty}^{\infty} dy' S(x',y') P(x-x',y-y'):
\label{eq:convolution}
\end{equation}
the predicted SQUID susceptibilities for the larger holes are the convolution of the point spread function $P(x-x',y-y')$ taken from Eq. \ref{eq:finalphi} with $S(x,y)$, the shape function for the hole. For example the shape function for a square hole of side $s$ would be $S(x,y)=1$ for $|x|<s/2, |y|<s/2$, $0$ otherwise.

Comparison of Figs. \ref{fig:multiple_experiment} and \ref{fig:multiple_theory} shows progressively poorer agreement as the holes get larger, for several reasons. First, the theory assumes circular symmetry, which is clearly not the case for the susceptometer diagrammed in Fig. \ref{fig:multiple_experiment}(a). Second, the implicit assumption of a linear, local response in Eq. \ref{eq:convolution} ignores the non-local nature of the response of a superconductor to applied magnetic fields.

One way to address the issue of the symmetry of the sensor is to deconvolute the point spread function from experimental data \cite{Roth1989using} using 
\begin{equation}
\bar{P}(k_x,k_y)=\bar{\phi}(k_x,k_y)*H(k,k_{\rm max})/\bar{S}(k_x,k_y)
\label{eq:psf}
\end{equation}
where $\bar{P}(k_x,k_y)$, $\bar{\phi}(k_x,k_y)$, and $\bar{S}(k_x,k_y)$ are the Fourier transforms of the point spread function $P(x,y)$, the measured susceptibility $\phi(x,y)$, and the hole shape $S(x,y)$, respectively. The Hanning function, $H(k,k_{\rm max})=(1+\cos(k/k_{\rm max}))/2$ for $k=\sqrt{k_x^2+k_y^2}<k_{\rm max}$ and 0 otherwise, was used to limit high frequency noise in $k$-space, with $k_{\max}$ = 12 $\mu$m$^{-1}$. The predictions of this approach have the asymmetry expected from the susceptometer layout, but ``fill in" more rapidly with hole size than experiment. We note that this second approach, while it improves the accuracy of the susceptometer shape, does not address the non-local nature of the response of a superconductor to applied magnetic fields.

A more complete modeling of defects with finite size would involve a numerical solution of London's and Maxwell's equations using a realistic geometry for the susceptometer and sample.\cite{KirtleyRevSciInstrum16,KirtleySupercondSciTechnol16}

\subsection{\label{naturally occurring}Naturally occurring defects}
\begin{figure}
\centering
\includegraphics[width=\linewidth]{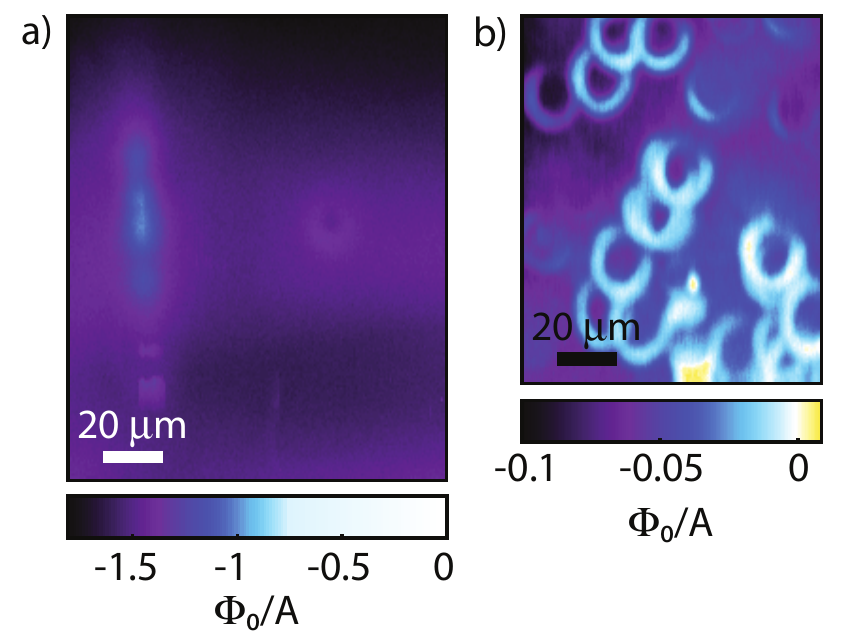}
\caption{{\bf We observe ``halos," approximately circular features of reduced susceptibility, in several two-dimensional superconductors}. In-phase susceptibility images of a) $\delta$-doped SrTiO$_{3}$ with $5.5$ nm thick, 1 at.\% Nb doping layer (studied in detail in Ref. \onlinecite{NoadPRB16}) and a set temperature of 150 mK; and b) LaAlO$_{3}$/SrTiO$_{3}$ (Sample G of Ref.~\onlinecite[Chapter 3]{NoadThesis17}), where the temperature as measured at the sample thermometer was 54 mK before and 84 mK after the scan.}
\label{fig:examples}
\end{figure}
\begin{figure}
\centering
\includegraphics[width=\linewidth]{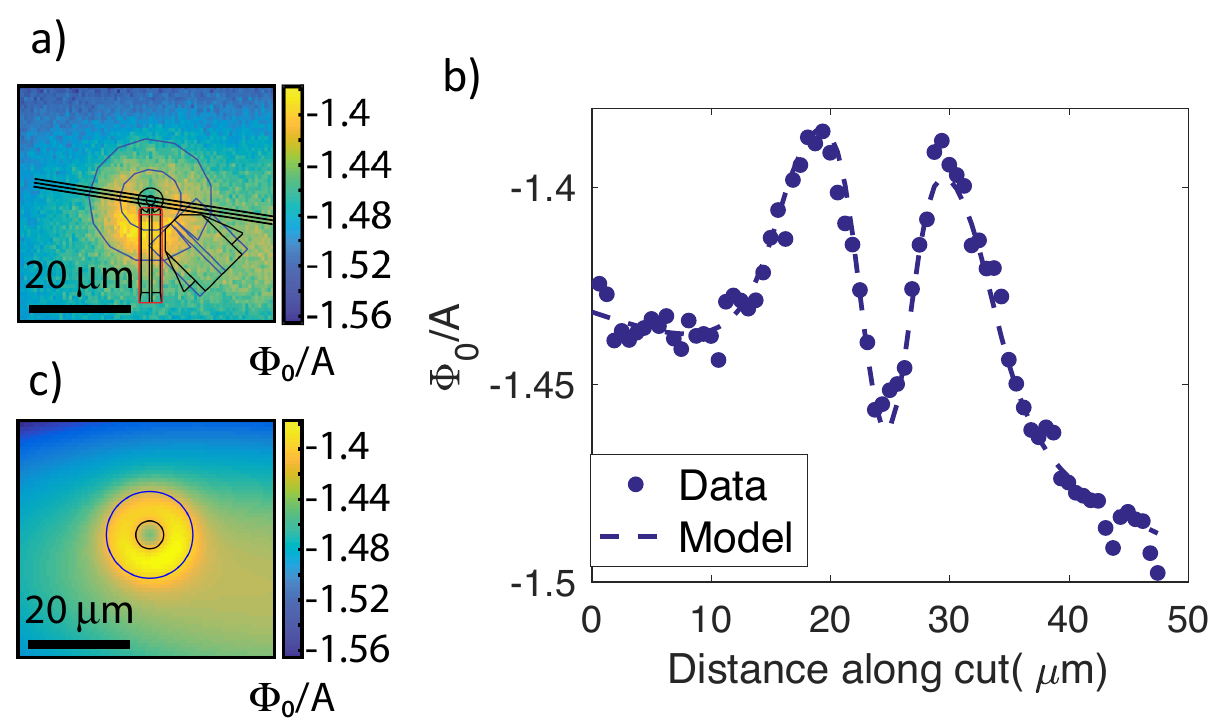}
\caption{{\bf The profile of a halo along a high-symmetry direction is reproduced by our model for SQUID susceptometry of a point-like defect.} a) In-phase susceptibility image taken, with field coil current 0.25 mA$_{rms}$ at 1472 Hz, on 5.5 nm, 1 at.\% Nb $\delta$-doped STO (cropped version of Fig.~\ref{fig:examples}(a)). Superimposed on the image are the positions of three line cuts, and the layout of the pickup loop (black) and field coil (blue)  region of the susceptometer  used for this image. \cite{HuberRevSciInstrum08} b) Average of line-cuts (dots)
and fit to (Eq.~\ref{eq:finalphi}) plus linear background (dashed line). c) Simulated image calculated with same parameters as in (b) and a second-order background determined by fitting a surface to the data far from the circular feature in (a). The field coil and pickup loop (blue and black overlays, respectively) are represented by concentric, co-planar circles of 8.4 $\mu$m and 2.7 $\mu$m radii, respectively.}
\label{fig:comparison}
\end{figure}
\begin{figure*}
\centering
\includegraphics[width=\linewidth]{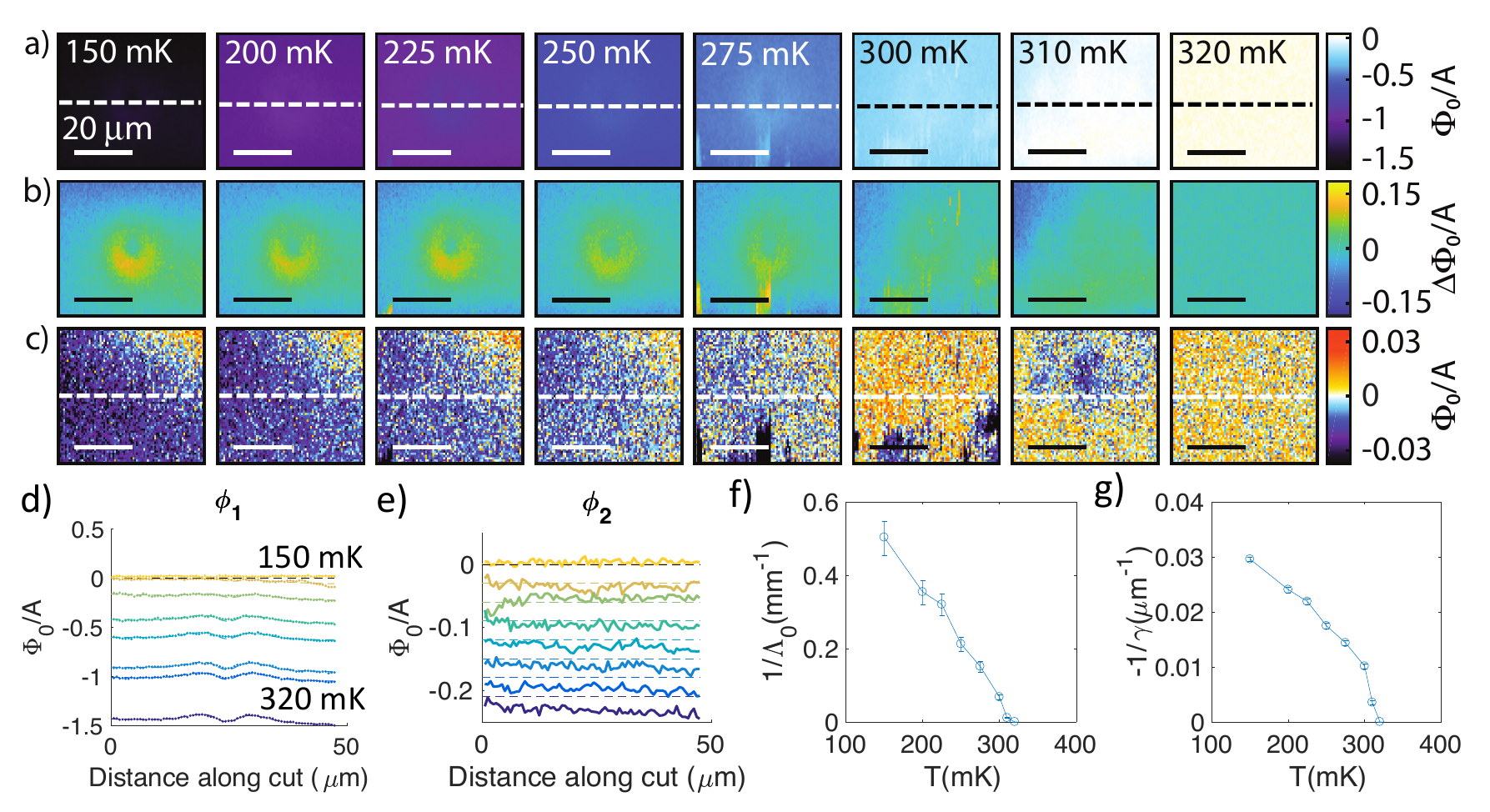}
\caption{{\bf Temperature dependence of defect image in $\delta$-doped STO}. Images of in-phase susceptibility a) without and b) with background subtraction and c) out-of-phase susceptibility without background subtraction in the region of the $\delta$-doped STO sample plotted in Fig.~\ref{fig:comparison}, as a function of temperature. Temperatures noted on the figure were measured at the mixing chamber at the beginning 
of each scan. Images and associated measurements of susceptibility as a function of height were taken at a field coil current of 0.25 mA$_{rms}$, 1472 Hz. d) Line-cuts through the in-phase ($\phi_1$) images as indicated by the dashed lines in a). e) Line-cuts through the out-of-phase ($\phi_2$) images as indicated by the dashed lines in c). The zeros of each curve (dashed lines) are offset by -0.02$\Phi_0/A$ as the temperature is lowered. f) and g) best fit values for the Pearl length $\Lambda_0$ and the defect strength parameter $\gamma$ from fits to the in-phase susceptibility $\phi_1$ as described in the text.}
\label{fig:tser_dSTO}
\end{figure*}
We observe halos of reduced diamagnetic susceptibility in both two-dimensional superconducting systems considered here, as shown in Fig.~\ref{fig:examples}(a, b). We have also observed similar features in a bulk superconductor.\cite{unpub_bulkLBCO} The diameter of the halos in the two-dimensional samples is similar to the $\sim 20~\mu$m diameter of the field coil in the SQUID susceptometer used. 
Our experience is that in LAO/STO samples there were areas with many haloes, but it was also possible to find areas with few or no haloes, whereas in the $\delta$-doped STO, it was rare to find halos at all and even more rare to find more than one in a single scan area. Further, it was possible to produce haloes in LAO/STO samples by repeatedly touching the sample with the susceptometer substrate.


Fig.~\ref{fig:comparison}(a) is an enlargement of the halo feature located in the center-right of Fig.~\ref{fig:examples}(a). Averaged linecuts perpendicular to the axis of bilateral symmetry of the halo show a double-peaked feature, Fig.~\ref{fig:comparison}(b), on top of an approximately linear background. The dashed curve in this figure is obtained by fitting the data to Eq.~\ref{eq:finalphi} plus a linear background of unknown origin, plus an offset flux $\phi_0$ given by the expression for the SQUID susceptibility of a thin film diamagnet\cite{KirtleyPRB12}
\begin{equation}
\phi_0 = -a\phi_s(1-2{\bar z}/\sqrt{1+4{\bar z}^2})/\Lambda_0
\label{eq:phi0}
\end{equation}
where $\Lambda_0$ is the Pearl length away from the defect, ${\bar z}=z_0/a$, and $\phi_s$ is the mutual inductance between the field coil and the pickup loop in the absence of a sample. This fit gave values for $\Lambda_0$=1.98$\pm$0.19 mm, and $\gamma$ = -33.7$\pm$0.5$\mu$m,
using fixed values of $a$ = 8.4~$\mu$m, $b$ = 2.7~$\mu$m,\cite{KirtleyPRB12} 
and $z_0$ = 2.9$\pm$0.3 $\mu$m, calculated using the scan offset height set in the measurement, the known sensor geometry, and an estimate for the SQUID-sample surface angle. The errors in the fit values for $\Lambda_0$ are dominated by the uncertainty in $z_0$, whereas the errors in $\gamma$ are dominated by statistical errors. Since for a homogeneous thin film and in the absence of fluctuations the Pearl length $\Lambda_0=2\lambda_0^2/d$, where $\lambda_0$ is the London penetration depth and $d$ is the film thickness, and the 2D superfluid density $N_s$ is given by $N_s = md/\mu_0 q^2 \lambda_0^2$, where $m$ is the mass and $q$ is the charge of the superconducting charge carriers, it follows that $1/\Lambda_0(T)$ is a measure of the 2D superfluid density $N_s(T)$. A Pearl length of 1 mm corresponds to a 2D superfluid density of $N_s=3\times10^{12}$ 1/cm$^2$. The extraordinarily long Pearl lengths and $\gamma$  values reported in this paper are a result of the very low 2D superfluid densities in LAO/STO and $\delta$-doped STO.

Fig.~\ref{fig:comparison}(c) shows a simulated image of a halo calculated using the same parameters as in (c) plus this same background.

\begin{figure*}
\centering
\includegraphics[width=\linewidth]{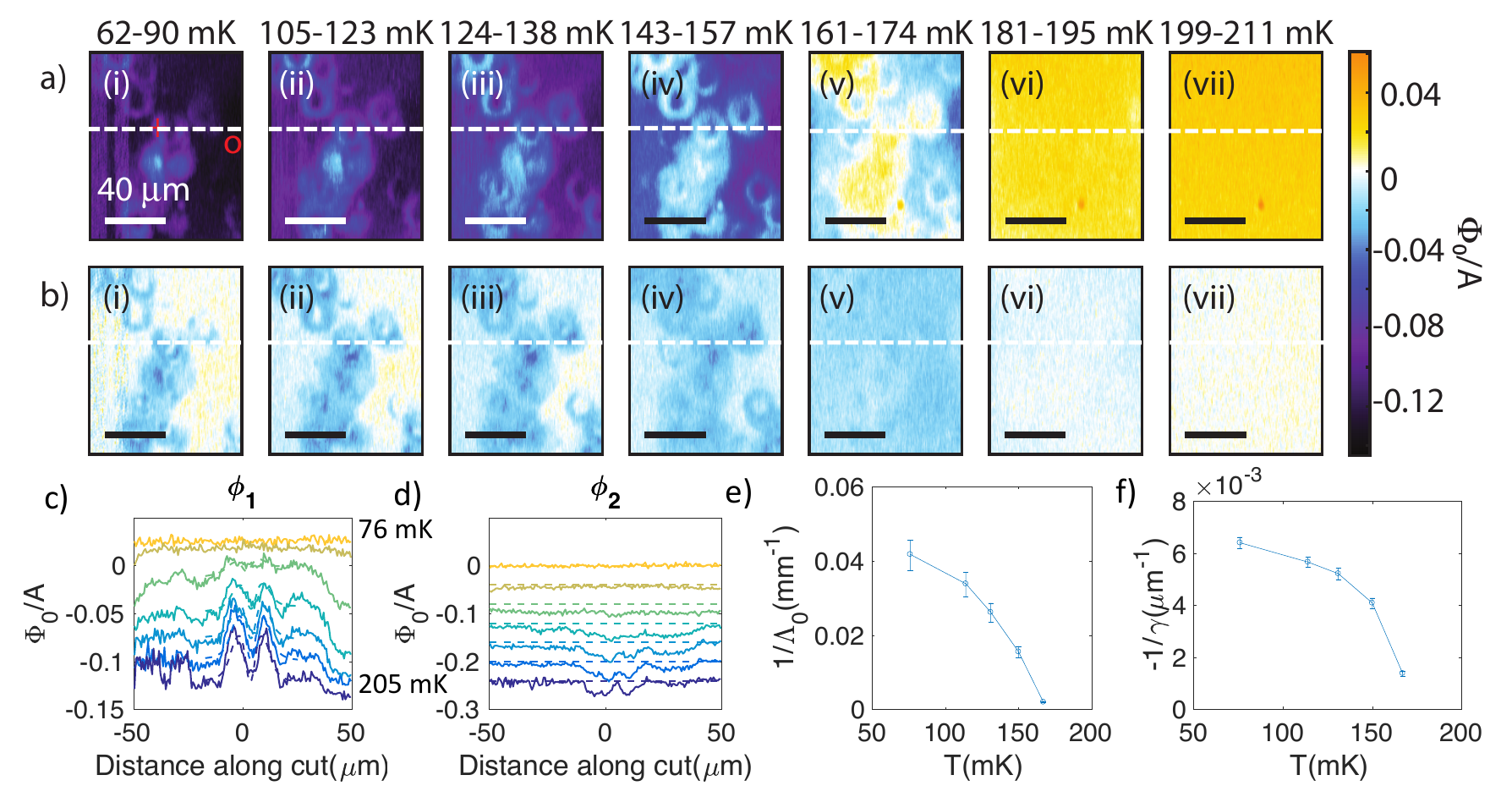}
\caption{{\bf Temperature dependence of defect images in LAO/STO.} Images of a) in-phase and b) out-of-phase susceptibility taken as a function of temperature on the same 5 u.c. LAO/STO sample as the data shown in Fig.~\ref{fig:examples}(b). The labels at the top of the images display the temperatures at the mixing chamber at the beginning and end of the scans. During this temperature series, a backgate voltage of 0 V was applied to the LAO/STO. Images and associated susceptibility versus height measurements taken at field coil current of 1 mA$_{rms}$, $1863.3$ Hz. b) Line-cuts through the in-phase susceptibility images ($\phi_1$) as indicated by the dashed lines in Fig. \ref{fig:tser_LAOSTO}a. d) Line-cuts through the out-of-phase images ($\phi_2$) as indicated by the dashed lines in \ref{fig:tser_LAOSTO}b. e) and f) best fits values for the uniform Pearl length $\Lambda_0$ and defect strength parameter $\gamma$ from fits to the in-phase susceptibility as described in the text.
}
\label{fig:tser_LAOSTO}
\end{figure*}

Figure \ref{fig:tser_dSTO} shows the evolution with temperature of the in-phase susceptibility $\phi_1$ without (Fig. \ref{fig:tser_dSTO}a) and with (Fig. \ref{fig:tser_dSTO}b) background subtraction, and the out-of-phase susceptibility $\phi_2$ (Fig. \ref{fig:tser_dSTO}c) of the halo of Fig.~\ref{fig:comparison}(a) at a low field coil current ($0.25$ mA$_{rms}$). Fig. \ref{fig:tser_dSTO}d,e displays horizontal cross-sections through the images as indicated by the dashed lines in Figures \ref{fig:tser_dSTO}a and \ref{fig:tser_dSTO}c respectively. As the temperature is lowered, the superfluid density becomes larger, the Pearl length becomes shorter, and the background susceptibility becomes more diamagnetic, as expected from Eq. \ref{eq:phi0}. Fits to this data using the same procedure as for Fig. \ref{fig:comparison},  displayed as dashed lines in Fig. \ref{fig:tser_dSTO}d, result in values for $1/\Lambda_0(T)$ and $1/\gamma(T)$ that decrease monotonically from low $T$ to $T_c$. Fig.~\ref{fig:tser_dSTO}c and Fig.~\ref{fig:tser_dSTO}e reveals no measurable out-of-phase halo signature even as the temperature approaches $T_{c}$ near 320 mK. In contrast, measuring a different sample (LAO/STO) as a function of temperature at 1 mA$_{rms}$, Fig.~\ref{fig:tser_LAOSTO}, reveals features in the out-of-phase susceptibility [Fig.~\ref{fig:tser_LAOSTO}(b)] at locations corresponding to halos in the in-phase channel [Fig.~\ref{fig:tser_LAOSTO}(a)], even at $T/T_{c} \sim 0.3$ [Fig.~\ref{fig:tser_LAOSTO}(b)(i)]. 

In LAO/STO, the susceptibility data on the halos and in the background region have different temperature dependences. Looking at a point away from any halos, such as the one indicated in Fig.~\ref{fig:tser_LAOSTO}(a)(i) at center right by a red `$\circ$', we see at the lowest temperature the strongest diamagnetic signal in the in-phase channel [Fig.~\ref{fig:tser_LAOSTO}(a)(i)] and no signal in the out-of-phase channel [Fig.~\ref{fig:tser_LAOSTO}(b)(i)].  As the temperature is steadily increased, the strength of the diamagnetism is gradually weakened [Fig.~\ref{fig:tser_LAOSTO}(a)(ii-v)] before becoming paramagnetic in the normal state [Fig.~\ref{fig:tser_LAOSTO}(a)(vi-vii)].  In the out-of-phase channel, the signal in the background region remains small and mostly featureless as the temperature is increased [Fig.~\ref{fig:tser_LAOSTO}(b)(ii-iii)] until just below $T_c$, where it peaks [Fig.~\ref{fig:tser_LAOSTO}(b)(iv-v)].  In the normal state, the out-of-phase component returns to zero [Fig.~\ref{fig:tser_LAOSTO}(b)(vi-vii)].

Considering now a point on a halo, such as indicated by the red `+' in Fig.~\ref{fig:tser_LAOSTO}(a)(i), the in-phase signal is already less diamagnetic than the background even at the lowest temperature [Fig.~\ref{fig:tser_LAOSTO}(a)(i)] and has a net paramagnetic response at a lower temperature [Fig.~\ref{fig:tser_LAOSTO}(a)(v)] than the surrounding background region, consistent with there being a lower $T_c$ associated with the halos than for the background region.  Likewise, the enhanced out-of-phase susceptibility of the `+' pixel is already visible at the lowest temperature [Fig.~\ref{fig:tser_LAOSTO}(b)(i)] and persists over a wider range of temperatures [Fig.~\ref{fig:tser_LAOSTO}(b)(ii-iv)] than the surrounding region (`$\circ$' symbol), as seen by faint lightening in the halo regions in   Fig.~\ref{fig:tser_LAOSTO}(b)(v).  By the time the surrounding region is normal, as indicated by a non-diamagnetic background, the out-of-phase signal associated with the halos has already also vanished, leaving Fig.~\ref{fig:tser_LAOSTO}(b)(vi-vii) featureless.  That is, in addition to a shift in temperature associated with the lower $T_c$ in halo regions, we also see the peak in the out-of-phase component broadened so much that it persists to our lowest measured temperatures. These trends are reproduced in the line-cuts displayed in Fig.'s \ref{fig:tser_LAOSTO}c,d.


\begin{figure*}
\centering
\includegraphics[width=\linewidth]{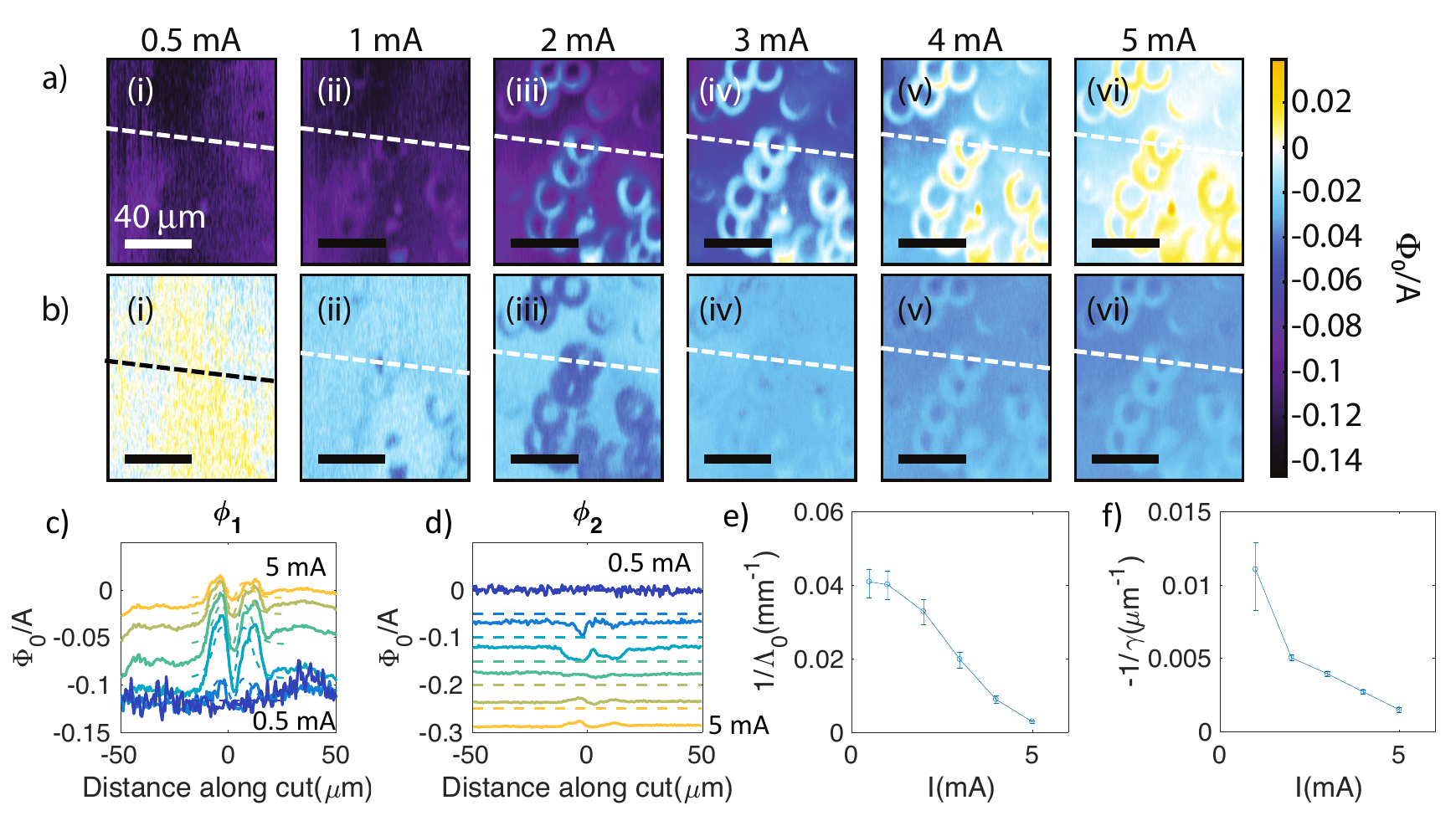}
\caption{{\bf Field coil current dependence of defect images in LAO/STO.} Susceptibility images as a function of field coil current at temperatures of $\sim 0.4 - 0.5$ $T/T_{c}$ on the same 5 u.c. LAO/STO sample as the data shown in Fig.~\ref{fig:examples}(b) and Fig.~\ref{fig:tser_LAOSTO}. a) in-phase and b) out-of-phase susceptibility. During this field coil series, a backgate voltage of -30 V was applied to the LAO/STO. RMS field coil currents as noted on images; excitation frequency of 1863.3 Hz for all images and associated measurements of susceptibility versus height. c) Line-cuts through the in-phase susceptibility ($\phi_1$) images as indicated by the dashed lines in a). d) Line-cuts through the out-of-phase susceptibility ($\phi_2$) images as indicated by the dashed lines in b). Successive curves are offset by -0.05$\Phi_0/A$ as the field coil current is increased. e) and f) Best fit values for the uniform Pearl length $\Lambda_0$ and defect strength parameter $\gamma$ as a function of field coil current $I$, obtained by fitting the in-phase susceptibility of (c).}
\label{fig:FCser_LAOSTO} 
\end{figure*}


The susceptibility of the LAO/STO measured as a function of field coil current (equivalently, applied field: this  field coil, with effective radius 8.4 $\mu$m, applies approximately 75 $\mu$T/mA to the sample) shows similar trends as in the temperature series [Fig.~\ref{fig:FCser_LAOSTO}].  At the lowest field coil current, both channels are nearly featureless [Fig.~\ref{fig:FCser_LAOSTO}(a, b)(i)].  Away from the halos, the in-phase component gradually approaches but does not reach zero as the current amplitude is increased [Fig.~\ref{fig:FCser_LAOSTO}(a)(ii-vi)].  The out-of-phase component away from the halos steadily increases but does not clearly reach its peak in the measured range [Fig.~\ref{fig:FCser_LAOSTO}(b)(ii-vi)].  By contrast, the in-phase susceptibility on halos crosses zero [Fig.~\ref{fig:FCser_LAOSTO}(a)(iv)] and shows a net paramagnetic signal at higher field amplitudes [Fig.~\ref{fig:FCser_LAOSTO}(a)(v-vi)] as seen in the normal state in the temperature series [Fig.~\ref{fig:tser_LAOSTO}(a)(vii)].  Furthermore, the out-of-phase component does peak on halos [Fig.~\ref{fig:FCser_LAOSTO}(b)(iii)], falling off as the field amplitude is further increased [Fig.~\ref{fig:FCser_LAOSTO}(b)(iv-vi)]. The sign of the out-of-phase component of the haloes relative to the background reverses at sufficiently high fields. A full theory of the out-of-phase component of scanning SQUID microscopy will be required to explain some of these results.


\section{Discussion}

\subsection{Comparison to model}
While the calculated curves in Fig.~\ref{fig:comparison}(c) and Fig.~\ref{fig:multiple_theory}(a) are in good agreement with the data, there are some notable differences between our images of halos and the calculated images. 

First, in all of the experimental data, we see gaps in each halo, whereas the calculated signals are circularly symmetric. The gap is an artifact from the physical layout in our sensors: the shields of the pickup loop (red in Fig. \ref{fig:comparison}(a) and Fig. \ref{fig:multiple_experiment}(a)) and the field coil (black in Fig. \ref{fig:comparison}(a) and Fig. \ref{fig:multiple_experiment}(a)) reduce the field applied to the sample when the sensor is positioned higher in $y$ than the defect in Fig. \ref{fig:comparison}(a) or lower in $y$ than the defect in Fig. \ref{fig:multiple_experiment}(a). 

Second, the experimental signal on the $\delta$-doped sample [Figs.~\ref{fig:examples}(a), \ref{fig:comparison}(a), \ref{fig:tser_dSTO}(a)] appears to fade towards the top of the image. This slope in the signal is present in the unperturbed signal far from the halo as well, and may either be a true change in the strength of the susceptibility signal with position or an artifact due to poor scan plane compensation. The calculated image in Fig.~\ref{fig:comparison} includes a second-order background slope determined from the data around the edge of the scan area, away from the halo, and it shows a similar fading-out of the halo towards the top of the image. 

Finally, the halos in Figs.~\ref{fig:examples}(b), \ref{fig:tser_LAOSTO}, and \ref{fig:FCser_LAOSTO} are somewhat stretched along the vertical and/or compressed along the horizontal, compared to the expected circular shape. This distortion is almost certainly an artifact of imperfect spatial calibration of the piezoelectric scanners.

Overall, the model confirms our interpretation of the halos in susceptibility as originating from enhancements in $\Lambda$ (reductions in $n_s$) at regions whose maximum spatial extent is smaller than our sensor. In principle, if the defect is actually point-like, the model allows us to pinpoint the position of the defect to better than the diameter of the SQUID pickup loop; the defect should be centered on the local minimum between the two lobes of a cross-section of a halo. 

All other parameters being equal, $\gamma$ tells us the relative strength of a sub-resolution defect. If, for example, a scan area contained several non-overlapping halos, we could compare $\gamma$ from halo to halo and perhaps infer something about the nature of the defects. The model does not capture any information about the structure of the defect; the defect is simply a delta function with zero spatial extent and perfect rotational symmetry. Physical defects are not delta functions, of course. The effective ``strength" of a defect that we might extract from images will depend not only on the actual local change in $\Lambda$, but also on the shape and spatial extent of the defect.

For the data shown in Fig.~\ref{fig:comparison}, we obtain a value of $\gamma = -21~\mu$m. To determine the size of the Pearl length (Eq. \ref{eq:GammaLambda}) in the defect from $\gamma$, one would have to assume an effective area for the defect. Supposing that the defect were just smaller than the pickup loop, with a radius of 1.5 $\mu$m, for example, $\gamma = -21~\mu$m would imply a perturbed $\Lambda$ of $2.3$ mm.

\section{Conclusion}
We have observed halo-like features in susceptibility images from intentionally introduced holes in superconducting Nb films as well as from two two-dimensional oxide superconductor system. A straightforward model confirms our interpretation of the halo features as originating from regions of increased $\Lambda$ (decreased $n_s$) on length scales smaller than those of our SQUID susceptometer. This understanding expands our toolkit for characterizing superconducting thin films.

It would be interesting to compare estimates of vortex pinning potentials, such as those given by vortex dragging experiments \cite{StraverAPL08, AuslaenderNatPhys09} or by studying the statistics of the positions of vortices over many field-cooling cycles, to the defect strength, $\gamma$, that we can extract from halos. We expect stronger suppressions of $n_s$ to correspond to stronger pinning potentials.

While the model that we used to calculate a halo can tell us where the defects are located, it cannot tell us about the structure or composition of the defects by itself. Complementary measurements, such as scanned Laue microscopy and strain mapping,\cite{microdiff1, microdiff2, microdiff3} would enable us to access such information.

\begin{acknowledgements}
This work was supported by the Department of Energy, Office of Science, Basic Energy Sciences, Materials Sciences and Engineering Division, under Contract DE-AC02-76SF00515. H. N. acknowledges support from a Stanford Graduate Fellowship and a Natural Sciences and Engineering Council of Canada PGS M and PGS D. We would like to thank Masayuki Hosoda for assistance with sample fabrication and Vladimir Kogan for useful discussions. We acknowledge Elie Track, Micah Stoutimore and Vladimir Talanov of Northrop Grumman Mission Systems for fruitful discussions and for providing samples for this work.
\end{acknowledgements}

\bibliography{references}

\end{document}